\documentclass{aa}
\usepackage{graphicx}                                      
\usepackage{txfonts}
\newcommand{\be}{\begin{equation}}
\newcommand{\ee}{\end{equation}}

\begin{document}

\title{Three-fluid plasmas in star formation \\
II. Momentum transfer rate coefficients}
\titlerunning{Momentum transfer rate coefficients}
\author{Cecilia Pinto$^1$, Daniele Galli$^2$}  
\authorrunning{Pinto \& Galli}
\institute{
$^1$Dipartimento di Astronomia e Scienza dello Spazio,
Universit\`a di Firenze,
Largo E. Fermi 5, I-50125 Firenze, Italy 
\\
$^2$ INAF--Osservatorio Astrofisico di Arcetri,
Largo E. Fermi 5, I-50125 Firenze, Italy 
\\}
\offprints{C. Pinto}
\date{Received date; accepted date}

\abstract
{The charged component of the insterstellar medium consists of atomic
and molecular ions, electrons, and charged dust grains, coupled to the
local Galactic magnetic field.  Collisions between neutral particles
(mostly atomic or molecular hydrogen) and charged species, and between
the charged species themselves, affect the magnetohydrodynamical
behaviour of the medium and the dissipation of electric currents.}
%
{The friction force due to elastic collisions between particles of
different species in the multi-component interstellar plasma is a
nonlinear function of the temperature of each species and the Mach
number of the relative drift velocity. The aim of this paper is to provide
an accurate and, as far as possible, complete set of momentum transfer
rate coefficients for magnetohydrodynamical studies of the interstellar
medium.}
%
{Momentum transfer rates are derived from available experimental data
and theoretical calculations of cross sections within the classic
approach developed by Boltzmann and Langevin for a wide range of values
of the temperature and the drift velocity.}
%
{Accurate numerical values for momentum transfer rates are obtained and
fitted to simple analytical formulae expressing the dependence of the
results on the gas temperature and the relative drift velocity.
The often used polarization approximation is in satisfactory agreement
with our results only for collisions between H$_2$ and molecular ions (HCO$^+$, H$_3^+$). For other kinds of collisions, the polarization
approximation fails by large factors, and must be replaced by
more accurate expressions.}
%
{}

\keywords{Atomic processes;Molecular processes;Plasmas;MHD;ISM:clouds;ISM:jets and outflows;ISM:magnetic fields}

\maketitle

\section{Introduction}

The interstellar medium (ISM) is a multi-component plasma that consists
mostly of hydrogen, helium, heavy ions, electrons and charged dust
grains.  The interaction between these components, and their coupling
with the Galactic magnetic field, determine the dynamical properties of
the ISM, control its evolution and the nature of the star formation
process.  In particular, the momentum exchange in collisions between
neutral and charged particles is responsible for transfering the
effects of electric and magnetic forces to the neutral component,
allowing the magnetic field to drift out of weakly-ionized molecular
clouds (Mestel \& Spitzer~1956), damping the propagation of Alfv\`en
waves (Zweibel \& Josafatsson~1983), and heating the gas by the
frictional dissipation of turbulent energy (Scalo 1977).

In a companion paper (Pinto, Galli \& Bacciotti~2007, hereafter
Paper~I), we have derived the equations governing the dynamics of a
three-fluid system, reducing the set of equations to a momentum
equation for the mean fluid and an evolution equation for the magnetic
field, plus two relations for the drift velocities in terms of the mean
fluid velocity and the magnetic field.

In this paper, we report on a detailed analysis of collisional rate
coefficients involving the most abundant neutral and charged species in
the ISM. The paper is organized as follows: in Sect.~\ref{sec_friction},
we give the general expression for the friction force and the momentum
transfer rate coefficient for elastic collisions, and we obtain an
analytical solution for a cross section varying as a power of the
relative velocity; in Sect.~\ref{sec_h2coll}, we consider collisions
with H$_2$ of HCO$^+$, H$_3^+$, H$^+$, and electrons, using available
theoretical and/or experimental determination of the collision cross
section; similarly, in Sect.~\ref{sec_hcoll}, we consider collisions
with H of C$^+$, H$^+$, and electrons; in Sect.~\ref{sec_hecoll}, we consider collisions of H$^+$ and electrons with He; in Sects.~\ref{sec_graincoll} and
\ref{sec_chargecoll} we consider collisions between charged dust grains
and neutral particles, and between charged particles, respectively; in
Sect.~\ref{sec_anal}, we give analytical approximations for our
numerical results; finally, in Sect.~\ref{sec_concl}, we summarize our
conclusions.

\section{Friction force and rate coefficients}
\label{sec_friction}

The general expression of the momentum acquired per unit time and unit
volume (``friction force'') by a particle of species $s$ with mass
$m_s$ and initial velocity ${\bf v}_s$ (``test
particle'') through collisions with particles of species $s^\prime$
with mass $m_{s^\prime}$ and initial velocity ${\bf v}_{s^\prime}$
(``field particles'') was given by Boltzmann~(1896),
\begin{eqnarray}
\lefteqn{
{\bf F}_{ss^\prime}=n_s n_{s^\prime}\int d{\bf v}_{s}f({\bf v}_s)
\int d{\bf v}_{s^\prime}f({\bf v}_{s^\prime})v_{ss^\prime}
}\nonumber \\
& & \times \int d\Omega {d\sigma\over d\Omega} m_s({\bf w}_{s}-{\bf v}_{s}),
\label{fcoll1}
\end{eqnarray}
where $f({\bf v}_s)$ and $f({\bf v}_{s^\prime})$ are the velocity
distribution functions of the two species, ${\bf w}_s$ is the velocity
of the test particle after the collision, $d\sigma/d\Omega$ is the
differential scattering cross section, and $v_{ss^\prime}$ is the
relative velocity of the particles (before the collision),
\be
v_{ss^\prime}\equiv |{\bf v}_{s}-{\bf v}_{s^\prime}|.
\ee

For elastic collisions, the last term in Eq.~(\ref{fcoll1}),
representing the momentum change of the test particle after the
collision, can be written as
\be
\int d\Omega{d\sigma\over d\Omega}m_s({\bf w}_s-{\bf v}_s)
=\mu_{s s^\prime}\sigma_{\rm mt}({\bf v}_{s^\prime}-{\bf v}_s),
\ee
where $\mu_{ss^\prime}=m_s m_{s^\prime}/(m_s+m_{s^\prime})$
is the reduced mass of the system,
\be
\sigma_{\rm mt}=\int {d\sigma\over d\Omega}(1-\cos\Theta)\,d\Omega
\label{sigmadiff}
\ee
is the {\rm momentum transfer} cross section, and $\Theta$ is 
{\it scattering angle} in the center-of-mass system.
Eq.~(\ref{fcoll1}) reduces then to the expression
\be
{\bf F}_{ss^\prime}=\mu_{ss^\prime}n_s n_{s^\prime}
\int d{\bf v}_{s}f({\bf v}_s)
\int d{\bf v}_{s^\prime}f({\bf v}_{s^\prime})v_{ss^\prime}
\sigma_{\rm mt}({\bf v}_{s^\prime}-{\bf v}_s).
\label{fcoll2}
\ee

The six integrations of eq.~(\ref{fcoll2}) require some care because of
the presence of the relative velocity in the integrand.  When the
velocity distribution of both species is maxwellian with temperatures
$T_s$ and $T_{s^\prime}$, the integrations can be carried out
explicitly, as shown first by Langevin~(1905), and the result is
\be
{\bf F}_{ss^\prime}= \alpha_{ss^\prime}({\bf u}_{s^\prime}-{\bf u}_s),
\label{fcoll3}
\ee
where
${\bf u}_s$ and ${\bf u}_{s^\prime}$ are the mean (maxwellian)
velocities of each species and the friction coefficient
$\alpha_{ss^\prime}$ is defined as
\be
\alpha_{ss^\prime}\equiv
\mu_{ss^\prime}n_sn_{s^\prime} \langle\sigma v\rangle_{ss^\prime}.
\label{frict}
\ee
In this expression, 
$\langle\sigma v\rangle_{ss^\prime}$ is an average of
the momentum transfer cross section over the relative velocity of the
interacting particles,
\begin{eqnarray}
\lefteqn{
\langle\sigma v \rangle_{ss^\prime} = 
a_{ss^\prime} {e^{-\xi^2}\over 2\sqrt{\pi}\xi^3}
}
\nonumber \\
& & \times \int_0^\infty \!\!
x^2 e^{-x^2}[(2\xi x-1)e^{2\xi x}+(2\xi x+1)e^{-2\xi x}]\sigma_{\rm mt} \,dx,
\label{integral}
\end{eqnarray}
where the nondimensional variables $\xi$ and $x$ are defined by
\be
\xi\equiv {v_d\over a_{ss^\prime}},\qquad 
x\equiv {v_{ss^\prime}\over a_{ss^\prime}},
\ee
and 
\be
v_d\equiv |{\bf u}_s-{\bf u}_{s^\prime}|
\ee
is the drift velocity, and
\be
a_{ss^\prime}\equiv\left({2kT_{ss^\prime}\over\mu_{ss^\prime}}\right)^{1/2},
\label{defa}
\ee
is the most probable velocity in a gas of particles of mass
$\mu_{ss^\prime}$ and temperature
\be
T_{ss^\prime} \equiv {m_s T_{s^\prime}+m_{s^\prime}T_s\over 
m_s+m_{s^\prime}}.
\ee
The quantity $\langle\sigma
v\rangle_{ss^\prime}$ is called the {\it momentum transfer rate
coefficient}. A related quantity is the {\it collision frequency}
between a test particle $s$ and the incident particles $s^\prime$
\be
\nu_{ss^\prime}=n_{s^\prime}\langle\sigma v\rangle_{ss^\prime}.
\ee
The momentum transfer rate coefficient defined by eq.~(\ref{integral})
is a function of the temperature $T_{ss^\prime}$ and the drift velocity
$|{\bf u}_{s^\prime}-{\bf u}_{s}|$, with the parameters $x$ and $\xi$
playing the roles of Mach number for the collision and the drift
velocity, respectively.  Owing to the dependence of $\langle \sigma
v\rangle_{ss^\prime}$ on $\xi$, the friction force eq.~(\ref{fcoll3})
is in general a nonlinear function of the drift velocity.

\begin{figure}
\resizebox{\hsize}{!}{\includegraphics{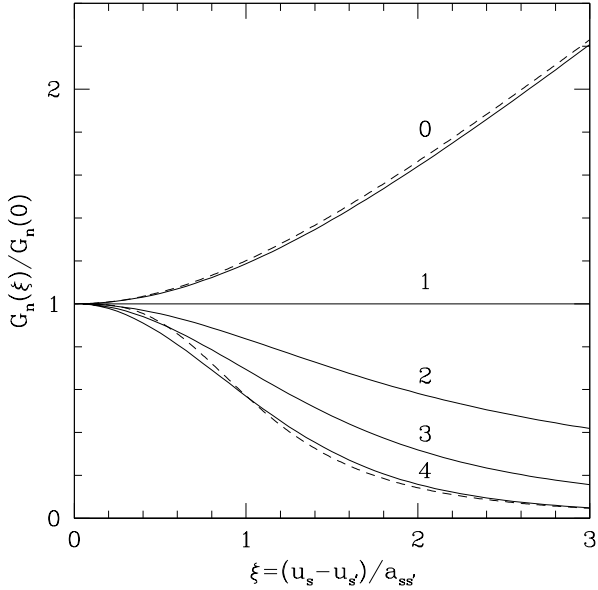}}
\caption{The normalized function $G_n(\xi)/G_n(0)$ defined by
eq.~(\ref{G_n}) as function of the normalized drift velocity $\xi=|{\bf
u}_s-{\bf u}_{s^\prime}|/a_{ss^\prime}$ with $n=0$, 1, 2, 3, and 4
({\it solid} curves). The {\it dashed} curves show the approximations
given by Draine \& Salpeter~(1979) for the cases $n=0$ and $n=4$.}
\label{fig_friction}
\end{figure}

For a cross section varying as a power-law of the relative velocity,
$\sigma(v_{ss^\prime})=\sigma_0x^{-n}$, the integral in
eq.~(\ref{integral}) can be expressed in terms of the confluent
hypergeometric function $M(a,b,c)$ (Abramowitz \& Stegun~1965),
\be
\langle\sigma v\rangle_n= \sigma_0 a_{ss^\prime} G_n(\xi), 
\label{sigmav_n}
\ee
where 
\be
G_n(\xi)=
\frac{4}{3\sqrt{\pi}}\Gamma\left(3-\frac{n}{2}\right) 
\exp(-\xi^2) M\left(3-\frac{n}{2},\frac{5}{2},\xi^2\right).
\label{G_n}
\ee
For vanishing drift velocities $G_n(\xi)$ reduces to a constant factor,
$G_n(0)=4\Gamma(3-n/2)/(3\sqrt{\pi})$, whereas in the opposite limit of
large drift velocities $G_n(\xi)\approx \xi^{1-n}$.  The behaviour of
the function $G_n(\xi)$ for $n=0$,1,2,3 and 4 is shown in
Fig.~\ref{fig_friction}.

Three special cases of eq.~(\ref{sigmav_n}) are relevant for collisions between 
ISM particles:

({\it a}\/) $n=0$,  with
\be
G_0(\xi)={1\over\sqrt{\pi}}\left(1+{1\over 2\xi^2}\right)
e^{-\xi^2}+\left(\xi +{1\over\xi}
-{1\over 4\xi^3}\right) {\rm erf}(\xi);
\label{g0}
\ee
(Baines et al.~1965). A cross section $\sigma_0=\pi(r_s+r_{s^\prime})^2$,
independent on the relative velocity describes the collisions between solid 
particles, represented as ``hard spheres'' of radii $r_s$ and $r_{s^\prime}$. 

({\it b}\/) $n=1$, with $G_1(\xi)=1$. 
A cross section varying as the inverse of the relative velocity arises
for an induced dipole attraction, where the interaction potential is
proportional to the inverse fourth power of the distance (``polarization
potential'').  Particles obeying  this particular interaction law are
called ``Maxwell molecules''(Maxwell~1860a,b), and the corresponding
collisional rate coefficient ``Langevin rate'' (see Appendix~A). In this
polarization approximation, the momentum transfer rate is independent
of both temperature and drift velocity.

({\it c}\/) $n=4$, corresponding to a (screened) Coulomb potential, with
\be
G_4(\xi)=\frac{{\rm erf}(\xi)}{\xi^3}-\frac{2e^{-\xi^2}}{\sqrt{\pi}\xi^2}.
\label{g4}
\ee
This result was first derived by Chandrasekhar~(1943) in the context of
stellar dynamics and, independently, by Dreicer~(1959) and
Sivukhin~(1966) for a fully-ionized gas\footnote{Draine \&
Salpeter~(1979) give the following accurate approximations for $n=0$ and 
$n=4$:
\[
G_0(\xi)\approx \frac{8}{3\sqrt{\pi}}\left(1+\frac{9\pi}{64}\xi^2\right)^{1/2},
\qquad
G_4(\xi)\approx \left(\frac{3\sqrt{\pi}}{4}+\xi^3\right)^{-1}.
\]
}. 

In many astrophysical applications, the relative velocity of the
interacting particles is of the order of the sound speed in the gas (if
the temperatures of the two species are not too different), but the
drift velocity $|{\bf u}_{s^\prime}-{\bf u}_{s}|$ is usually much lower
than the sound speed. In such circumstances it is appropriate to expand
the integrand in eq.~(\ref{integral}) for small values of the quantity $\xi x$,
obtaining, to the lowest order in the expansion, a value of the momentum
transfer rate coefficient independent of the drift velocity,
\be
\langle\sigma v\rangle_{ss^\prime} \approx 
{4\over 3}\left(\frac{8kT_{ss^\prime}}{\pi\mu_{ss^\prime}}\right)^{1/2} 
\int_0^\infty x^5 e^{-x^2}\sigma_{\rm mt}(x)\,dx+{\cal O}(\xi^2).
\label{aver_v}
\ee
If the momentum transfer cross section is given in terms of the
collision energy in the center-of-mass system $E_{\rm cm}=
\mu_{ss^\prime}v^2_{ss^\prime}/2$, the rate coefficient in the small
drift limit can be written as
\be
\langle\sigma v \rangle_{ss^\prime} \approx
{2\over 3}\left(\frac{8kT_{ss^\prime}}{\pi\mu_{ss^\prime}}\right)^{1/2} 
\int_0^\infty z^2 e^{-z}\sigma_{\rm mt}(z)\,dz+{\cal O}(\xi^2),
\label{aver_E}
\ee
where $z\equiv E_{\rm cm}/kT_{ss^\prime}$, a frequently-adopted
expression (see e.g., Mitchner \& Kruger~1973).
\footnote{The momentum transfer cross section measured in laboratory
experiments is usually given as function of the energy in the
laboratory frame ($E_{\rm lab}$). Before using eq.~(\ref{aver_E}),
$E_{\rm lab}$ must be converted into the energy in the center-of-mass
frame according to the formula
\[
E_{\rm cm}=\frac{m_s}{m_s+m_{s^\prime}}E_{\rm lab}
+\frac{3m_{s^\prime}}{2(m_s+m_{s^\prime})}kT_s,
\]
where $s$ and $s^\prime$ indicate the target and incident particles,
respectively.}.

\section{Collisions with H$_2$}
\label{sec_h2coll}

Collisions between H$_2$ molecules and charged particles determine the
rate of diffusion of the interstellar magnetic field through the
dominantly neutral gas of cold molecular clouds (see Paper~I).  In
applications to molecular clouds (e.g., Nakano~1984, Mouschovias~1996),
the collision rate coefficient $\langle\sigma v\rangle_{s,{\rm H}_2}$
has usually been estimated with the polarization approximation (see
Appendix) for collisions with molecular ions, and, also sometimes, for
collisions with electrons (see discussion in Mouschovias~1996). For
collisions between H$_2$ and dust grains, the hard sphere model
(Sect.~2) has generally been assumed. We review below the validity of
these assumptions and compute accurate values for the collision rate
coefficients using available momentum transfer cross sections.

\subsection{HCO$^+$ -- H$_2$}

\begin{figure}
\resizebox{\hsize}{!}{\includegraphics{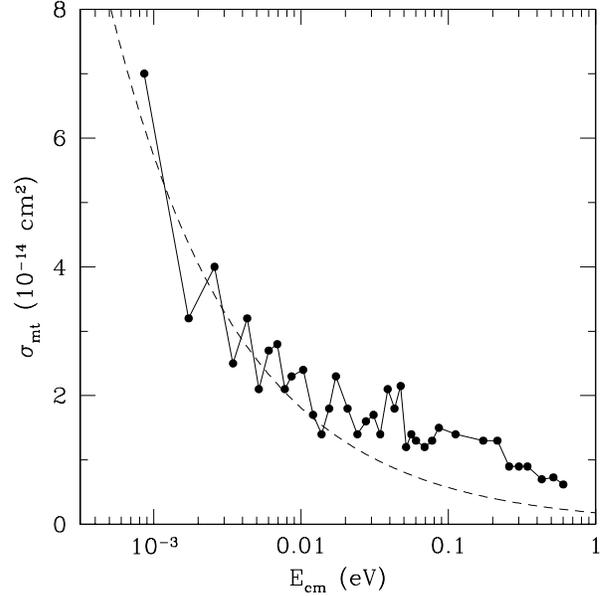}}
\caption{The momentum transfer cross section for collisions
HCO$^+$--H$_2$ as a function of $E_{\rm cm}$ computed by Flower~(2000)
({\it dots} and {\it solid} curve).  The {\it dashed} curve shows the
Langevin cross section.}
\label{hcoph2_cs}
\end{figure}

\begin{figure} \resizebox{\hsize}{!}{\includegraphics{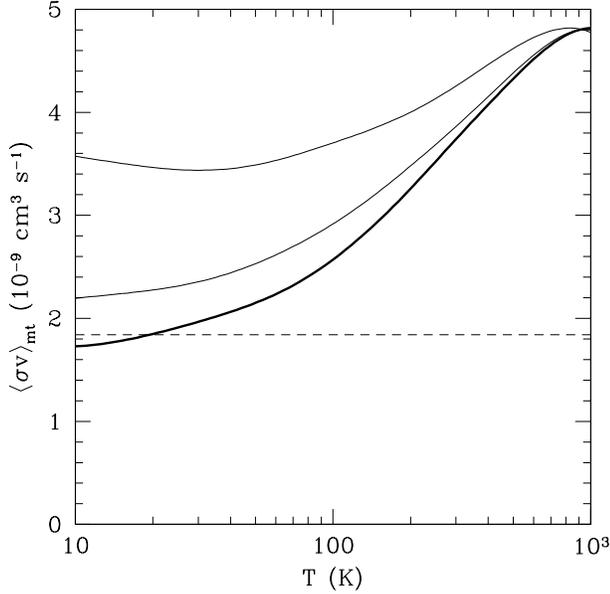}}
\caption{The rate coefficient for HCO$^+$--H$_2$ collisions as a function
of the temperature $T$ for $v_d=0$ ({\it thick solid} curve); and
$v_d=1$~km~s$^{-1}$; $v_d=2$~km~s$^{-1}$ ({\it thin solid} curves,
bottom to top), compared with the Langevin rate ({\it dashed} line).}
\label{hcoph2}
\end{figure}

Flower~(2000) has calculated quantum-mechanically the cross section
between HCO$^+$, a dominant ion in typical molecular cloud conditions,
and H$_2$ molecules in their rotational ground states.
Figure~\ref{hcoph2_cs} shows the cross section computed by Flower~(2000)
compared with the Langevin cross section.  As noticed by Flower~(2000),
the Langevin value gives a good approximation to the quantal results.
The rate coefficient is, therefore, expected to depend very weakly on
temperature and drift velocity, as shown in Fig.~(\ref{hcoph2})
\footnote{ Our rate coefficient for zero drift velocity differs by
$\sim 20$\% from the result by Flower~(2000) who adopts an averaging
over the particles kinetic energy slightly different from our
eq.~(\ref{aver_E}).}. Previous estimates of the molecular ion-H$_2$
rate coefficient based on the Langevin formula (see Fig.~\ref{hcoph2})
have ignored the weak dependence from temperature and drift velocity.
This neglect is, however, of little consequence for models of
magnetically-controlled cloud collapse as long as the temperature of
the infalling gas is in the range 10--20~K and the ion-neutral drift
velocity is a small fraction of the sound speed, as shown e.g., by
Mouschovias \& Ciolek~(1994).

\subsection{H$_3^+$ -- H$_2$}

\begin{figure}
\resizebox{\hsize}{!}{\includegraphics{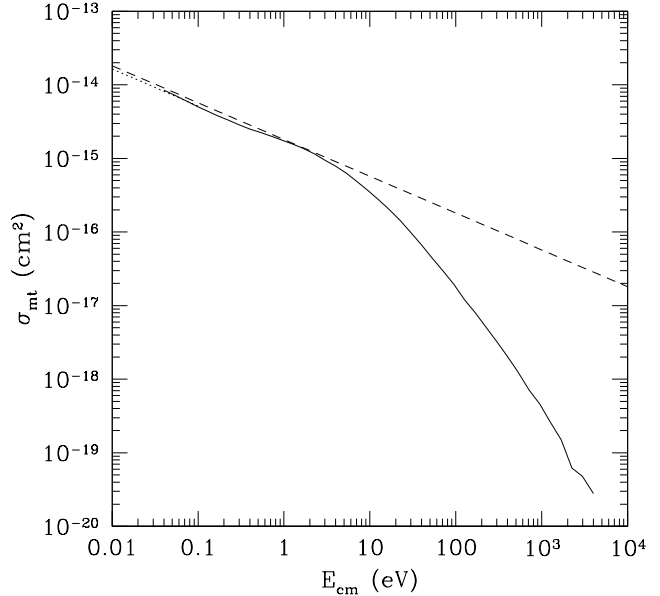}}
\caption{Momentum transfer cross section for H$_3^+$--H$_2$ collisions
as a function of the energy in the center-of-mass system according to
Phelps~(1990) ({\it solid} curve). The {\it dotted} line shows our
extrapolation at low energies, whereas the {\it dashed} line shows the
Langevin cross section.}
\label{h3ph2_cs}
\end{figure}

\begin{figure}
\resizebox{\hsize}{!}{\includegraphics{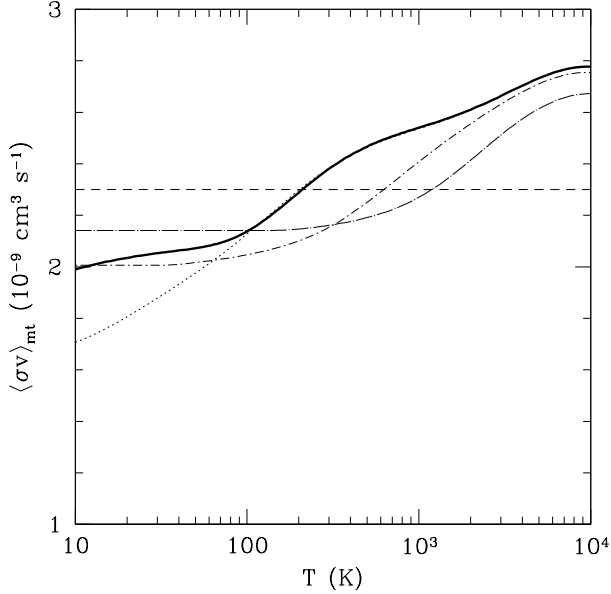}}
\caption{The momentum transfer rate coefficient for H$_3^+$--H$_2$
collisions as function of the temperature $T$ computed with the cross
section shown in Fig.~\ref{h3ph2_cs} for $v_d=0$ ({\it thick solid}
curve); $v_d=1$~km~s$^{-1}$ ({\it dotted}
curve); $v_d=5$~km~s$^{-1}$ ({\it short-dash dotted}
curve); and $v_d=10$~km~s$^{-1}$ ({\it long-dash dotted}
curve). The {\it dashed} line shows the Langevin rate.}
\label{h3ph2}
\end{figure}

Figure~\ref{h3ph2_cs} shows the momentum transfer cross section for
collisions of H$_3^+$ and H$_2$, recommended by Phelps~(1990), converted
to energies in the center-of-mass frame.  At low energies ($E_{\rm cm}
\lesssim 1$~eV), the cross section is obtained from the results of the
mobility experiments of Ellis et al.~(1976). At higher energies, the
behavior of the cross section is extrapolated from mobility data at
room temperature.  As shown in the figure, the cross section is very
close to the Langevin value up to $E_{\rm cm}\approx 3$~eV, but
declines steeply (as $E_{\rm cm}^{-1.7}$) above $E_{\rm cm}\approx
10$~eV. Since the cross section tabulated by Phelphs~(1990) does not
extend to ion energies below 0.1~eV, we have extrapolated Phelps's
value to lower energies with the asymptotic formula $\sigma_{\rm
mt}=1.62\times 10^{-15} (E_{\rm cm}/\mbox{eV})^{-0.507}$.
Figure~\ref{h3ph2} shows the corresponding rate coefficient obtained
integrating the cross section, according to Eq.~(\ref{integral}) as a
function of the temperature and for various values of the drift
velocity.  As in the case of HCO$^+$, given the Langevin behavior of
the cross section at low energies, the dependence of the collisional
rate from temperature and drift velocity is very weak for temperatures
below $\sim 10^4$~K.

\subsection{H$^+$ -- H$_2$}

\begin{figure}
\resizebox{\hsize}{!}{\includegraphics{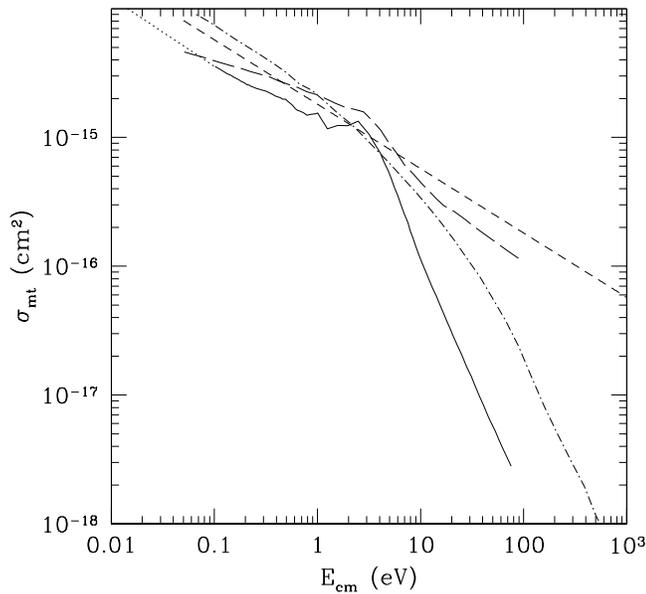}}
\caption{The momentum transfer cross section for H$^+$--H$_2$
collisions as a function of $E_{\rm cm}$:  fully-quantal calculation by
Krsti\'c \& Schultz~(1999) ({\it solid} curve), with our extrapolation to 
low energies ({\it dotted} line); classical calculation
by Bachmann \& Reiter (1995) ({\it long-dashed} curve); semi-empirical
recommendations by Phelps~(1990) ({\it dot-dashed} curve); Langevin
cross section ({\it dashed} curve).}
\label{hph2_cs}
\end{figure}

\begin{figure}
\resizebox{\hsize}{!}{\includegraphics{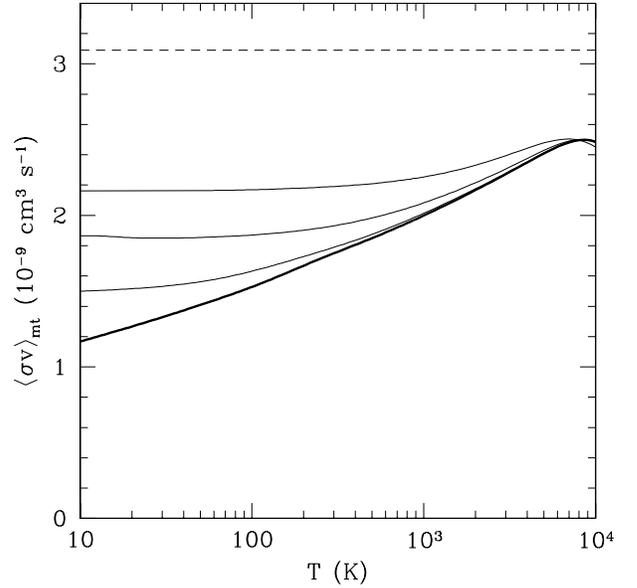}}
\caption{The collision rate coefficient for H$^+$--H$_2$ collisions as
a function of the temperature, computed with the momentum transfer cross
section of Krsti\'c \& Schultz~(1999) for $v_d=0$ ({\it thick solid}
curve); and $v_d=2$~km~s$^{-1}$; $v_d=4$~km~s$^{-1}$;
$v_d=10$~km~s$^{-1}$ ({\it thin solid} curves, bottom to top). The {\it
dashed} line shows the Langevin rate.}
\label{hph2}
\end{figure}

Figure~\ref{hph2_cs} shows the momentum transfer cross section for
H$^+$--H$_2$ collisions computed by Krsti\'c \& Schultz~(1999) in the
energy range $0.1~\mbox{eV}<E_{\rm cm}<100~\mbox{eV}$ (fully-quantal
calculation), by Bachmann \& Reiter~(1995) in the energy range
$0.05~\mbox{eV} < E_{\rm cm} < 10^2~\mbox{eV}$ (classical calculation)
and the cross section recommended by Phelps~(1990) in the energy range
$0.067~\mbox{eV}< E_{\rm cm}< 6.67~\mbox{keV}$ (obtained interpolating
results from ion mobility experiments for $E_{\rm cm}<2.3$~eV and
theoretical cross sections above 330~eV). As shown in
Fig.~\ref{hph2_cs}, the quantal results depart significantly from the
Langevin value at high energies ($E_{\rm cm}> 4$~eV), and differ
substancially from the results extrapolated from ion mobility
experiments in the same energy range. Here we adopt the cross section
computed by Krsti\'c \& Schultz~(1999), extrapolated to energies below
$0.1$~eV with the asymptotic formula $\sigma_{\rm mt}=1.05\times
10^{-15} (E_{\rm cm}/\mbox{eV})^{-0.531}$.  We show in Fig.~\ref{hph2}
the corresponding rate coefficient as a function of the temperature and
the drift velocity. Again, the rate coefficient depends very weakly on
these two quantities for temperatures below $\sim 10^4$~K.

\subsection{$e$ -- H$_2$}

The elastic scattering of electrons by H$_2$ has been the subject of
much theoretical and experimental work of the past decades.  In
general, it is well established that the momentum transfer cross
section for electron-molecule scattering deviates significantly from
the classical Langevin rate at low energies (below $\sim 1$~eV), owing
to the effect of ``electron exchange", i.e., the exchange of the
incoming electron with one orbital electron in the neutral species (see
e.g., Massey \& Ridley~1956, Morrison \& Lane~1975). The existence of
this effect has been confirmed experimentally by Ferch et al.~(1980) in
the energy range 0.02--2~eV.  The net result is a reduction of the
momentum transfer cross section, which compensates the polarization
(Langevin) contribution in such a way that the resulting cross section
is roughly constant at low-collision energies.

\begin{figure}
\resizebox{\hsize}{!}{\includegraphics{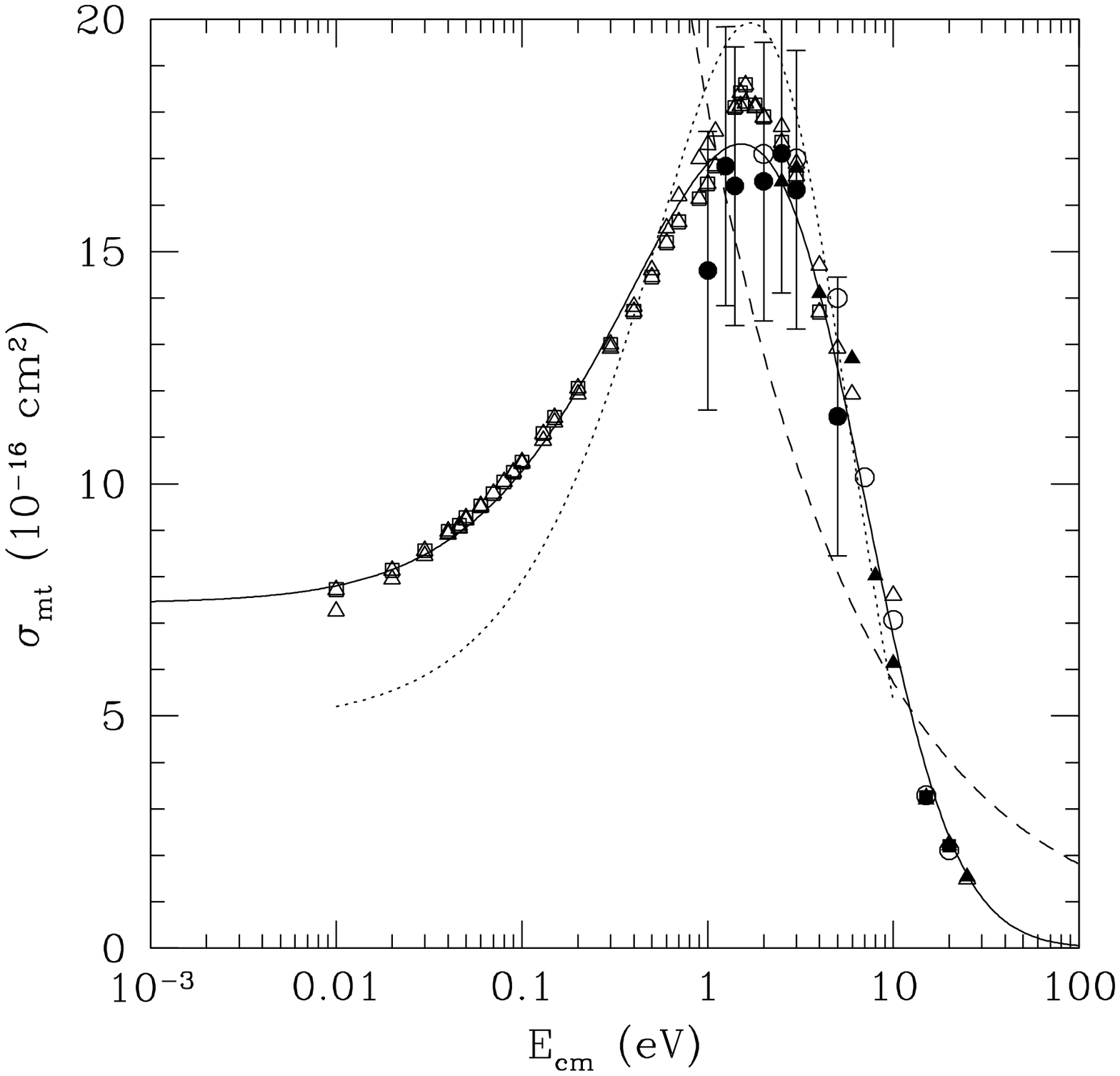}}
\caption{The momentum transfer cross section for $e$--H$_2$ collisions
as a function of the electron kinetic energy.  Experimental values:
England et al.~(1988) ({\it empty triangles}), Schmidt et al.~(1994)
({\it empty squares}), Shyn \& Sharp~(1981) ({\it empty circles}),
Nishimura, Danjo \& Sugahara~(1985) ({\it filled triangles}), Khakoo \&
Trajmar~(1986) ({\it filled squares}), Brunger et al. (1990,1991) ({\it
filled circles}).  The {\it dashed} and {\it dotted} curves show the
Langevin cross section and the quantum-mechanical theoretical results
of Henry \& Lane~(1969), respectively.  The {\it solid} curve shows the
cross section adopted in this work.}
\label{eh2_cs}
\end{figure}

\begin{figure}
\resizebox{\hsize}{!}{\includegraphics{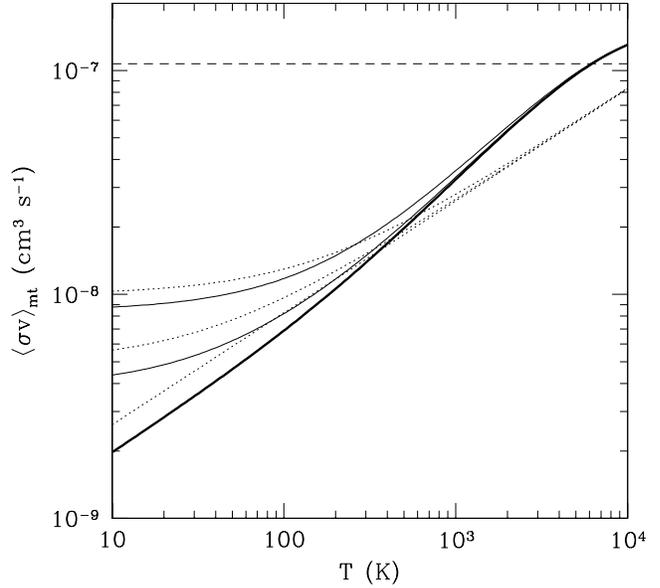}}
\caption{The momentum transfer rate coefficient for $e$--H$_2$
collisions computed with the cross section shown in Fig.\ref{eh2_cs} as a
function of the temperature $T$ for $v_d=0$ ({\it thick solid} curve);
and $v_d=50$~km~s$^{-1}$; $v_d=100$~km~s$^{-1}$ ({\it thin solid}
curves, top to bottom), compared with the values computed with Eq.~(14)
of Draine~(1983) for the same values of the drift velocity ({\it
dotted} curve).The {\it dashed} line shows the Langevin rate.}
\label{eh2}
\end{figure}

We have assembled a compilation of the available measurements of the
$e$-H$_2$ momentum transfer cross section, for collision energies
ranging from $10^{-3}$~eV up to $200$~eV. A detailed summary of the
most recent experimental results is given by Brunger \& Buckman~(2002).
The results are shown in Fig.~\ref{eh2_cs}, compared with the
theoretical calculations of Henry \& Lane~(1969) and the Langevin
value. Clearly, the semi-classical Langevin formula provides a poor
approximation to the actual cross section, especially at energies below
$\sim 1$~eV, where the effects of ``electron exchange'' are dominant.
The agreement between laboratory measurements and theoretical values of
the scattering cross sections appear to be satisfactory, with the
possible exception of the region of very low electron energies, where
the possible existence of a Ramsauer-Townsend minimum in the momentum
transfer cross section is not completely excluded (Ramanan \&
Freeman~1991).  The momentum transfer collision rate calculated with
the cross section indicated by the solid line in Fig.~\ref{eh2_cs} is
shown in Fig.~\ref{eh2}. At temperatures typical of interstellar
clouds, $T\approx 10$~K, the Langevin formula overestimates the actual
value of the momentum transfer rate by about two orders of magnitude.

\section{Collisions with H}
\label{sec_hcoll}

\subsection{C$^+$ -- H}

\begin{figure}
\resizebox{\hsize}{!}{\includegraphics{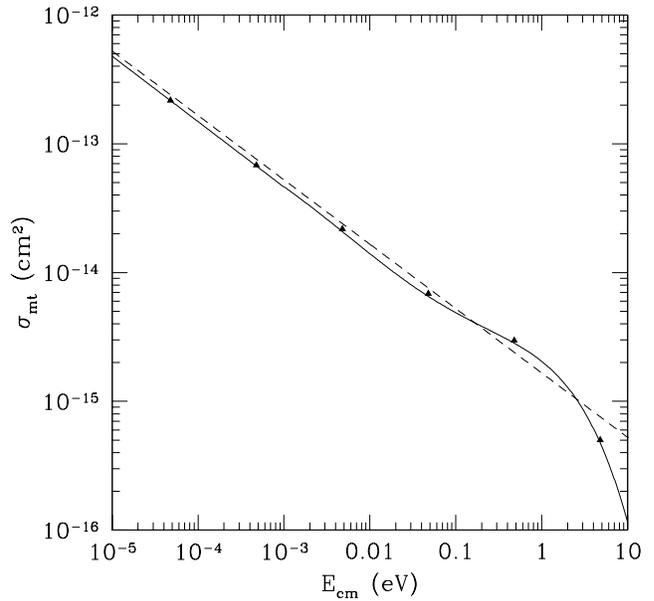}}
\caption[]{The momentum transfer cross section for collisions C$^+$--H
computed by Flower \& Pineau-des-F\^orets~(1995) as a function of the
collision energy in the center-of-mass frame (original data, {\it
triangles}; our interpolation, {\it solid} curve). The {\it dashed}
curve shows the Langevin cross section.}
\label{cph_cs}
\end{figure}

\begin{figure}
\resizebox{\hsize}{!}{\includegraphics{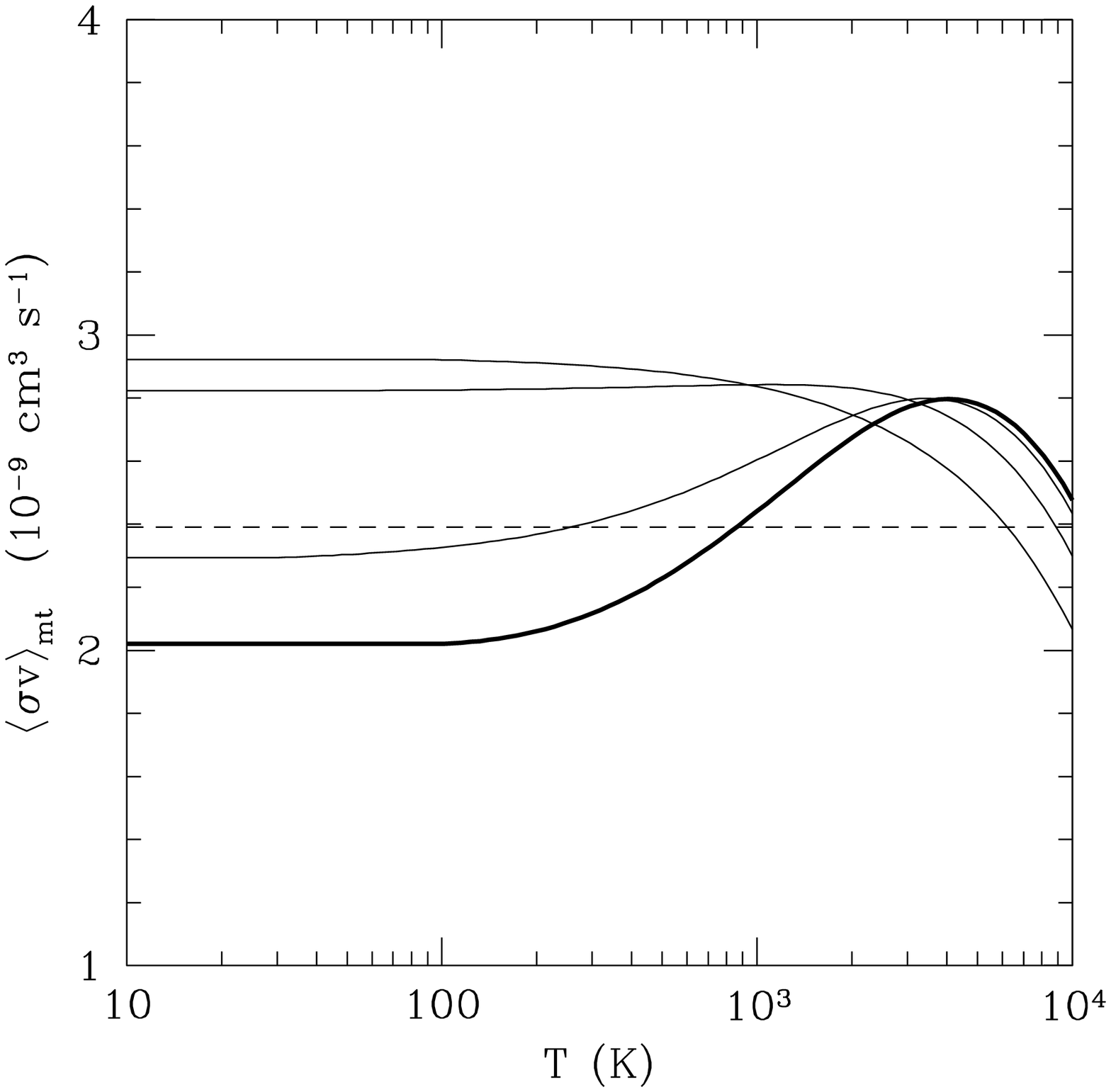}}
\caption[]{The collision rate coefficient for collisions C$^+$--H as a
function of the temperature for $v_d=0$ ({\it thick solid} curve); and
$v_d=5$; $v_d=10$; and $v_d=15$~km~s$^{-1}$ ({\it thin solid} curves, bottom to
top), compared with the Langevin rate ({\it dashed} line).}
\label{cph}
\end{figure}

The momentum transfer cross section for collisions of C$^+$ ions with H
atoms was computed by Flower \& Pineau-des-For\^ets~(1995)
at several values of the collision energy, ranging from $E_{\rm
cm}\approx 5\times 10^{-5}$~eV to $E_{\rm cm}\approx 5$~eV with the
adiabatic potential of Green et al.~(1972) (see Fig.~\ref{cph_cs}).
Figure~\ref{cph} shows the corresponding rate coefficient computed by
integrating numerically our interpolation of the results of Flower \&
Pineau-des-For\^ets~(1995).

\subsection{H$^+$ -- H}

\begin{figure}
\resizebox{\hsize}{!}{\includegraphics{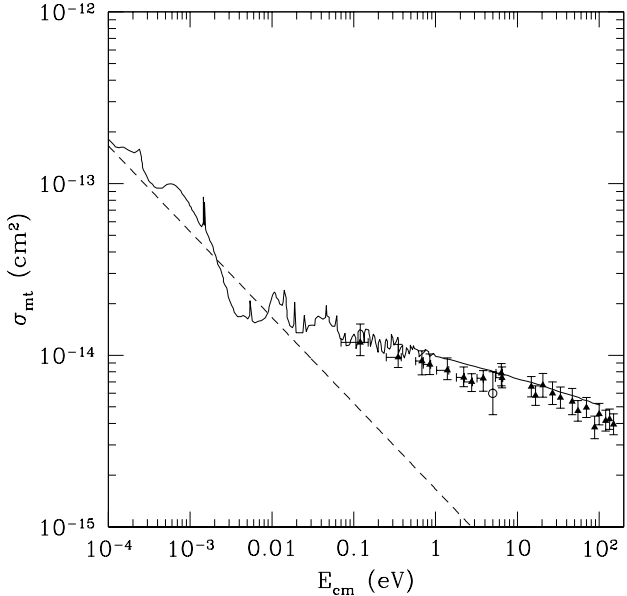}} \caption[]{The
momentum transfer cross section for collisions H$^+$--H computed by GKS
as a function of the collision energy in the center-of-mass frame ({\em
solid} curve), compared to the Langevin cross section ({\em dashed}
curve). The {\em empty circle} is the measurement by Brennan \&
Morrow~(1971) at $E_{\rm cm}\approx 5$~eV. The {\em filled triangles} are
the values of the charge exchange cross section for the same interacting
particles, measured by Newman et al.~(1982), multiplied by 2.}
\label{hp_cs}
\end{figure}

\begin{figure}
\resizebox{\hsize}{!}{\includegraphics{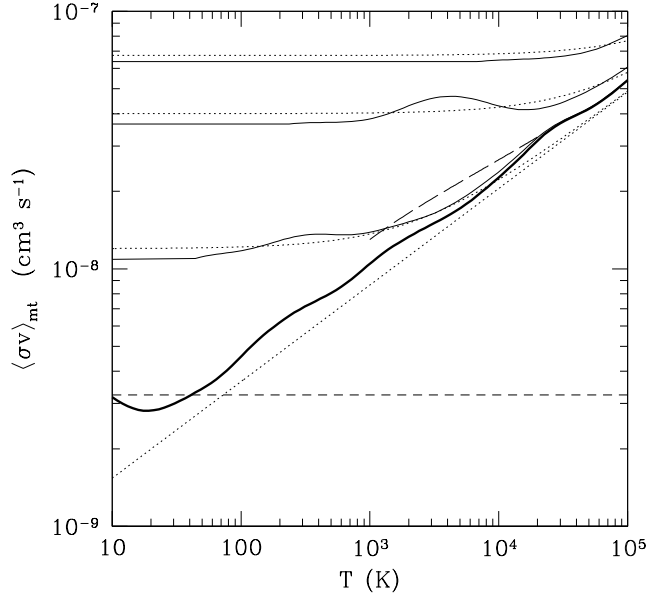}}
\caption[]{The collision rate coefficient for collisions H$^+$--H as a
function of the temperature for $v_d=0$ ({\em thick, solid} curve); $v_d=10$;
$v_d=50$~km~s$^{-1}$; and $v_d=100$~km~s$^{-1}$ ({\em thin, solid}
curves, bottom to top); compared with the fitting formula of GKS for the same values
of the drift velocity ({\em dotted} curves), the fitting formula 
of Geiss \& B\"urgi~(1986) for $v_d=0$ ({\em long-dashed} curve) and the Langevin value ({\em short-dashed} curves) .}
\label{hp}
\end{figure}

Being the simplest ion-neutral collision process, the scattering of
H$^+$ by H atoms has been the subject of a large number of theoretical
investigations, from the semiclassical calculations of Dalgarno \&
Yadav~(1953) to the accurate quantum mechanical calculations of
Krsti\'c \& Schultz~(1999) and Glassgold, Krsti\'c \& Schultz~(2005,
hereafter GKS). The latter paper, reporting a determination of the
momentum transfer cross section over a range from $10^{-10}$~eV to
$10^2$~eV (partially shown in Fig.~\ref{hp_cs}), represents the definitive
reference on this elastic process.  The only experimental determination
of the H--H$^+$ momentum transfer cross section was obtained by Brennan
\& Morrow~(1971) at $E_{\rm cm}\approx 5$~eV by observing the velocity
and attenuation of compressional Alfven waves propagating in a
partially-ionized hydrogen plasma. The measured value, $\sigma_{\rm
mt}=6.0^{+2.0}_{-1.5}\times 10^{-15}$~cm$^2$, is in good agreement with
the quantum mechanical results (see Fig.~\ref{hp_cs}).  Previous
estimates of the momentum transfer rate coefficient (Geiss \&
B\" urgi~1986) were based on measurements and calculations of the charge
exchange cross section $\sigma_{\rm ce}$ for the reaction ${\rm H}+{\rm
H}^+\rightarrow {\rm H}^++{\rm H}$ complemented at low energies with
the polarization cross section computed with the Langevin formula
eq.~(\ref{lange}). Since the charge transfer process proceeds with
little momentum transfer between the interacting particles,
eq.~(\ref{sigmadiff}) with $\Theta=\pi$ gives $\sigma_{\rm mt} \approx
2\sigma_{\rm ce}$ (Dalgarno~1958, Banks \& Holzer~1968).  As shown by
Fig.~\ref{hp_cs}, a combination of the polarization cross section and
twice the value of the charge exchange cross section roughly reproduces
the accurate results of GKS. However, the momentum transfer rate
estimated by Geiss \& B\"urgi~(1986) with this approximation for
temperatures between $10^3$~K and $2\times 10^4$~K is about 50\% higher
than the rate obtained by a numerical integration of the momentum
transfer cross section of GKS, shown in Fig.~\ref{hp} for different
values of the drift velocity.

\subsection{$e$ -- H}

\begin{figure}
\resizebox{\hsize}{!}{\includegraphics{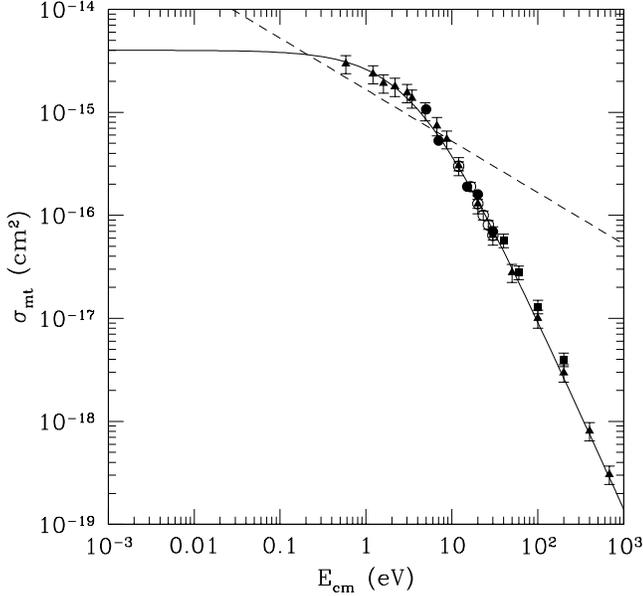}}
\caption[]{Momentum transfer cross section for $e$--H collisions.
Experimental values: Shyn \& Cho~(1989) ({\em filled circles});
Williams (1975a,b) ({\em filled triangles}); Shyn \& Grafe~(1992) ({\em
filled squares}); Callaway \& Williams~(1975) ({\em empty circles}).
The {\em solid} curve is the theoretical calculation of Dalgarno et
al.~(1999).}
\label{eh_cs}
\end{figure}

\begin{figure}
\resizebox{\hsize}{!}{\includegraphics{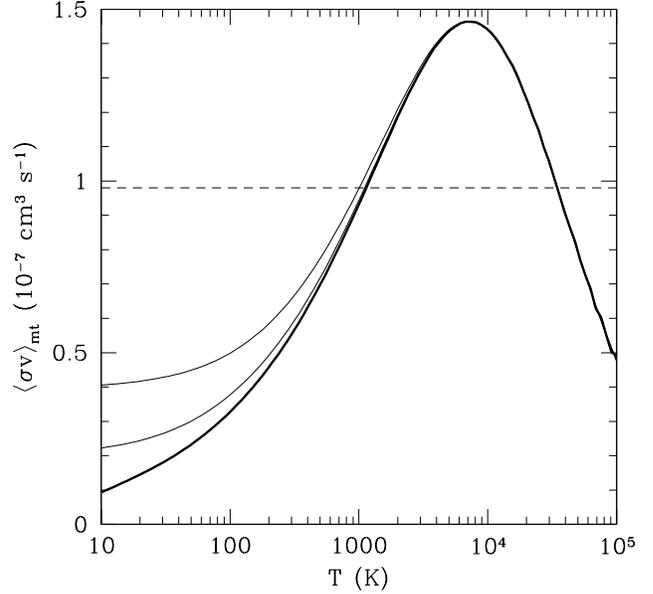}}
\caption[]{Collision rate coefficient for $e$--H collisions as a function
of temperature, for $v_d=0$~km~s$^{-1}$ ({\em thick} curve);
$v_d=50$~km~s$^{-1}$; and $v_d=100$~km~s$^{-1}$ ({\em thin} curves, bottom to top). The {\it dashed} line shows the Langevin rate.}
\label{eh}
\end{figure}

The momentum transfer cross section for $e$--H collisions has been
computed by Dalgarno, Yan, \& Liu~(1999) with the quantal formula of
Dalgarno \& Griffing~(1958) and the phase shifts of Rudge~(1975) and
Das \& Rudge~(1976) at $E \lesssim 10$~eV, and by van Wyngaarden \&
Walters~(1986) in the energy range 100--300~eV. Laboratory measurements
of the $e$--H momentum transfer cross sections have been performed by
Williams~(1975a,b), Callaway \& Williams~(1975), Shyn \& Cho~(1989),
and Shyn \& Grafe~(1992) (for a detailed review of the experimental
methods and results, see Bederson \& Kieffer~1971 and Trajmar \&
Kanik~1995). Theoretical and experimental results are shown in
Fig.~\ref{eh_cs}. The dependence of the cross section on collision
energy is clearly non-Langevin:  at low energies the cross section is
approximately constant, $\sigma\approx 4\times 10^{-15}$~cm$^{-2}$,
whereas at high energies, it decreases with energy as $E_{\rm
cm}^{-1.8}$. As a result, the momentum transfer rate, shown in
Fig.~\ref{eh}, is lower by about one order of magnitude than
the Langevin value at $T\approx 10$~K, and larger by $\sim 40\%$ at
$T\approx 10^4$~K, with a weak dependence on the relative drift velocity.

\section{Collisions with He}
\label{sec_hecoll}

With the exception of collisions with H$^+$ or electrons, to our knowledge,
there are no experimental or theoretical data available for elastic
collisions between He atoms and charged species. An estimate based on
the Langevin approximation (see Appendix), gives
\be
\langle \sigma v \rangle_{i,{\rm He}}= 
\left[ \frac{(m_i+m_{\rm He})m_n p_{\rm He}}{(m_i+m_n)m_{\rm He}p_n}
\right]^{1/2} \langle \sigma v \rangle_{in},
\label{he_lan}
\ee
where $n=$H or H$_2$, and $p_n$ is the polarizability 
of species $n$ (see Appendix).  According to
this equation, the momentum transfer rate for collision of ions with He
is, in general, a factor 0.4-0.5 of the corresponding rate for
collisions with H$_2$ and a factor 0.3-0.4 of the corresponding
rate for collisions with H.

Collisions with He introduce a small correction to the expression of the
friction coefficient $\alpha_{in}$ defined by eq.~(\ref{frict}). For a
neutral component made of H$_2$ and He, eq.~(\ref{he_lan}) gives
\be
\alpha_{in}\equiv \alpha_{i,{\rm H}_2}+\alpha_{i,{\rm He}}=
c_i\alpha_{i,{\rm H}_2},
\ee
with
\be
c_i=
1+\left[\frac{(m_i+m_{{\rm H}_2})m_{\rm He}p_{\rm He}}
{(m_i+m_{\rm He})m_{{\rm H}_2}p_{{\rm H}_2}}
\right]^{1/2} \left(\frac{n_{\rm He}}{n_{{\rm H}_2}}\right).
\ee
For example, for a cosmic He abundance of 0.1, the He correction
factors based on the Langevin approximation are $c_{{\rm H}^+}=1.12$,
$c_{{\rm H}_3^+}=1.13$, and $c_{{\rm HCO}^+}=1.15$ (see also
Mouschovias~1996). Similarly, for collisions with atomic hydrogen, the
He correction for the H--H$^+$ rate coefficient is $c_{{\rm
H}^+}=1.08$.

\subsection{H$^+$ -- He}

\begin{figure}
\resizebox{\hsize}{!}{\includegraphics{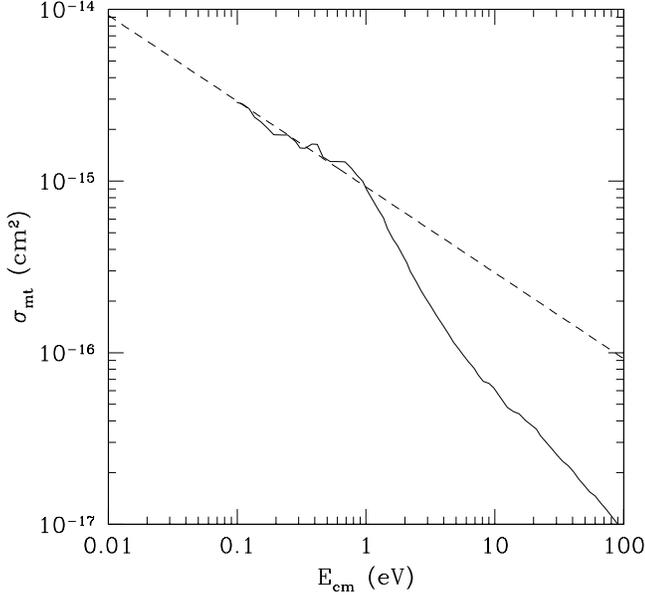}}
\caption[]{The momentum transfer cross section for collisions H$^+$--He,
according to the fully-quantal calculations of Krsti\'c \& Schultz~(1999),
as a function of the collision energy in the center-of-mass frame ({\em
solid} curve), compared to the Langevin cross section ({\em dashed}
line).}
\label{hep_cs}
\end{figure}

\begin{figure}
\resizebox{\hsize}{!}{\includegraphics{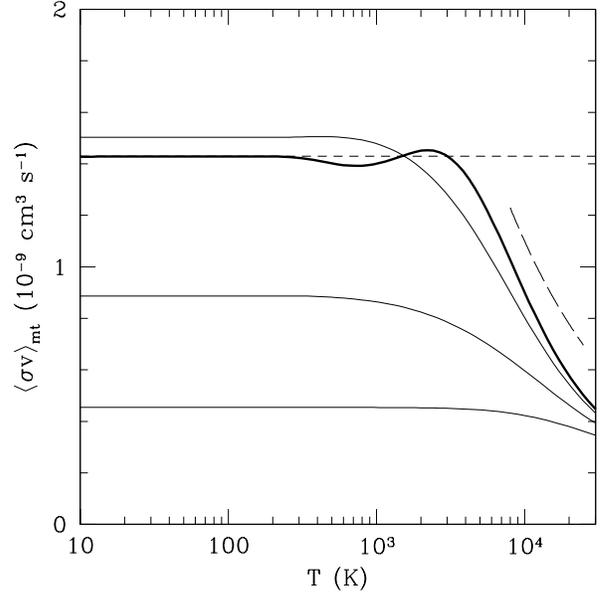}}
\caption[]{The collisional rate coefficient for H$^+$--He collisions as a
function of the temperature for $v_d=0$ ({\em thick, solid} curve);
$v_d=10$~km~s$^{-1}$; $v_d=20$~km~s$^{-1}$; and $v_d=30$~km~s$^{-1}$
({\em thin, solid} curves, top to bottom) compared with the fitting formula 
of Geiss \& B\"urgi~(1986) for $v_d=0$ ({\em long-dashed} curve) and the Langevin value ({\em short-dashed} curves).}
\label{hep}
\end{figure}

Figure~\ref{hep_cs} shows the momentum transfer cross section for
H$^+$--He collisions computed by Krsti\'c \& Schultz~(1999) with a semi-classical treatment in the energy range $0.1~\mbox{eV}<E_{\rm
cm}<100~\mbox{eV}$. The calculation has been recently extended up
to $10^4$~eV by Krsti\'c \& Schultz~(2006). The theoretical results are in
good agreement with the Langevin value below $E_{\rm cm} \approx 1$~eV,
but at larger energies the Langevin formula overestimates the theoretical
results (by one order of magnitude at $E_{\rm cm} \approx 100$~eV). To our
knowledge, no experimental results are available for H$^+$--He collisions.
Here we adopt the cross section computed by Krsti\'c \& Schultz~(1999),
extrapolated to energies below $0.1$~eV with the Langevin value. The
resulting collisional rate coefficient is shown in Fig.~\ref{hep} as a
function of the temperature and the drift velocity. As for H$^+$--H$_2$
collisions, the rate coefficient depends very weakly on these two
quantities for temperatures below $\sim 10^3$~K and drift speeds
below $\sim 10$~km~s$^{-1}$. The rate coefficient $\langle \sigma
v \rangle_{{\rm H}^+,{\rm He}}$ is about 0.9 $\langle \sigma v
\rangle_{{\rm H}^+,{\rm H}_2}$, whereas eq.~(\ref{he_lan}) gives a factor ~ 0.5.

\subsection{$e$ -- He}

Accurate values of the momentum transfer cross sections for $e$--He
collisions have been obtained from mobility experiments (Crompton, Elford \& Jory~1967;
Crompton, Elford \& Robertson~1970; Milloy \& Crompton~1977; Ramanan \&
Freeman~1990). The agreement between different experimental evaluations
is excellent, at the level of 1-2\%. As in the case of
$e$--H$_2$ collisions discussed in Sect.~\ref{sec_h2coll}, the actual
cross section deviates significantly from the classical Langevin value,
owing to the quantum exchange of the incoming electron with one orbital
electron of He. This is especially evident at low-collision energies,
where the momentum transfer cross section is approximately constant.
The rate coefficient $\langle \sigma v \rangle_{e,{\rm He}}$ is a factor $\sim 0.6$ of $\langle \sigma v \rangle_{e,{\rm H}_2}$.
\begin{figure}
\resizebox{\hsize}{!}{\includegraphics{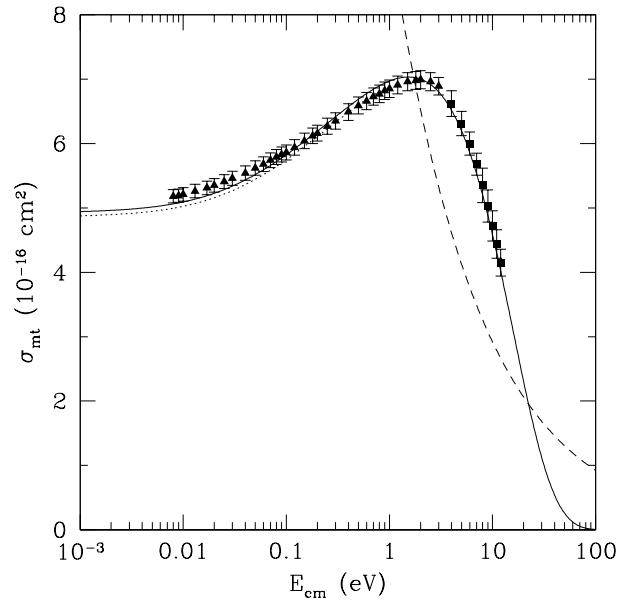}}
\caption{The momentum transfer cross section for $e$--He collisions as a
function of the electron kinetic energy.  Experimental
values: Crompton et al.~(1970) ({\it filled triangles}), Milloy \&
Crompton~(1977) ({\it filled squares}). The {\it dotted} curve shows
the experimental results of Ramanan \& Freeman~(1990), with uncertainty
$\sim 3$\%.  The {\it dashed} curve shows the Langevin cross section,
whereas the {\it solid} curve is the cross section adopted in this work.}
\label{ehe_cs}
\end{figure}

\begin{figure}
\resizebox{\hsize}{!}{\includegraphics{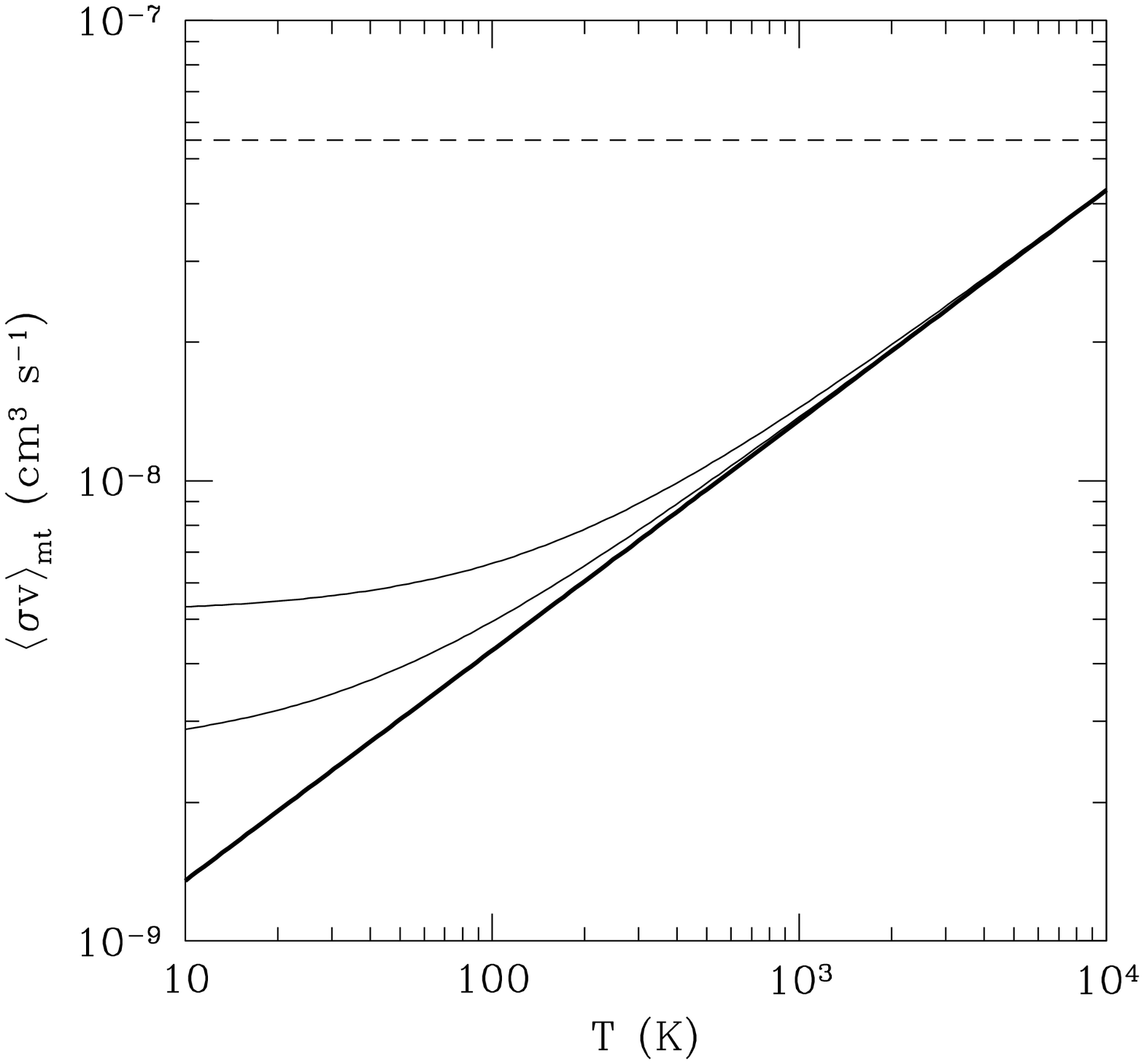}}
\caption{The momentum transfer rate coefficient for $e$--He
collisions computed with the cross section shown in Fig.\ref{ehe_cs} as a
function of the temperature $T$ for $v_d=0$ ({\it thick, solid} curve);
and $v_d=50$~km~s$^{-1}$; $v_d=100$~km~s$^{-1}$ ({\it thin, solid}
curves, bottom to top). The {\em dashed} line shows the Langevin rate.}
\label{ehe}
\end{figure}

\section{Collisions between dust grains and neutral particles}
\label{sec_graincoll}

For collisions between a spherical grain $g$ of charge $Z_g e$ and radius $r_g$ with 
a neutral particle 
$n$ of polarizability $\alpha_n$, the geometrical cross section $\pi r_g^2$ is 
larger that the polarization cross section (eq.~\ref{lange}) when
\be
r_g>24 |Z_g|^{1/4}\left(\frac{\alpha_n}{\mbox{\AA}^3}\right)^{1/4}
\left(\frac{T_{ss^\prime}}{\mbox{K}}\right)^{-1/4}~\mbox{\AA},
\ee
a condition generally satisfied by interstellar grains.
The average momentum transfer cross section for collisions of spherical 
grains with neutral atoms or molecules is then
\be
\langle\sigma v\rangle_{gn} = 
\pi r_g^2 \delta \left(\frac{2kT_{gn}}{\mu_{gn}}\right)^{1/2} G_0(\xi),
\label{hardspheres}
\ee
where $G_0(\xi)$ is given by eq.~(\ref{g0}) and 
$\delta$ is a factor of order unity (the so-called Epstein coefficient) equal 
to unity if the neutrals impinging on the grain undergo specular reflections.
For very subsonic drift velocity, this expression reduces to
\be
\langle \sigma v\rangle_{gn} \approx \frac{4}{3}\pi r_g^2 \delta
\left(\frac{8kT_n}{\pi m_n}\right)^{1/2},
\ee
a result derived by Epstein~(1924). Experiments with micron-size melamine-formaldehyde
spheres show that $\delta\approx 1.3$ (Liu et al.~2003). 

\begin{figure}
\resizebox{\hsize}{!}{\includegraphics{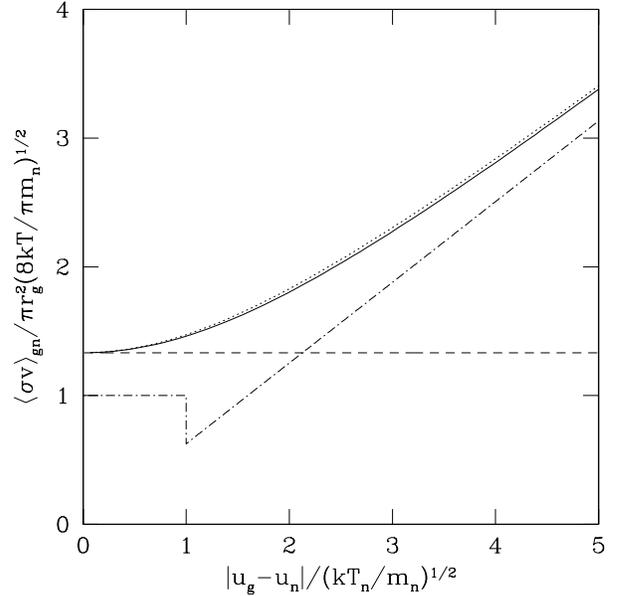}} 
\caption{Momentum transfer rate coefficient for collisions
grain--neutral as a function of the relative drift velocity, from
eq.~(\ref{G_n}) with $n=0$ ({\it solid} curve), compared with the
expressions from Draine \& Salpeter~(1979) ({\it dotted}
curve); Nakano~(1984) ({\it dashed} curve); and Mouschovias \& Ciolek
({\it dot-dashed} curve).}
\label{gn}
\end{figure}

Figure~\ref{gn} shows the grain-neutral momentum transfer rate as a
function of the relative drift velocity according to
eq.~(\ref{hardspheres}) with $\delta=1$ compared with the
approximations given by Draine \& Salpeter~(1979), Nakano~(1984), and
Mouschovias \& Ciolek~(1999). In this figure, the grain-neutral
momentum transfer rate is normalized to the expression given by
Mouschovias \& Ciolek~(1999) in the low-drift limit, and the drift
velocity is normalized to the mean thermal speed in the neutrals.

\section{Collisions between charged particles}
\label{sec_chargecoll}

Particles of charge $Z_{s^\prime} e$ can exchange momentum with
particles of charge $Z_s e$ and density $n_s$ via long-range Coulomb
interactions. In the standard Coulomb scattering theory (see e.g.,
Chapman \& Cowling~1953), the momentum transfer cross section is given
by
\be
\sigma_{\rm mt}=2\pi r_{\rm C}^2(v_{ss^\prime})
\ln\left[\frac{r_{\rm C}^2(v_{ss^\prime})+r_{\rm max}^2}
{r_{\rm C}^2(v_{gs})+r_{\rm min}^2}\right],
\ee
where 
\be
r_{\rm C}(v_{ss^\prime})=\frac{Z_sZ_{s^\prime}e^2}{\mu_{ss^\prime}v_{ss^\prime}^2}
\ee
is the Coulomb radius (the distance at which the electrostatic
energy is of the order of the kinetic energy) at the relative velocity
$v_{ss^\prime}$, and $r_{\rm min}$, $r_{\rm max}$ are the minimum and
maximum impact parameters of the collision. Several different
approaches have been taken to calculate $r_{\rm min}$ and $r_{\rm
max}$. For the latter, the standard choice is the total Debye length in
the plasma $\lambda_{\rm D}$ (Cohen et al.~1950), given by
\be
\frac{1}{\lambda_{\rm D}^2}=\sum_s\frac{1}{\lambda_{{\rm D},s}^2}=
\frac{4\pi e^2}{k}\sum_s\frac{Z_s^2n_s}{T_s}.
\ee

When averaging over the relative velocity of interacting particles
(eq.~\ref{integral}), the slowly varying logarithmic factor can be
considered constant, and the term $r_{\rm C}(v_{ss^\prime})$ in the
logarithm's argument can be replaced by
\be
\overline{r_{\rm C}}\equiv r_{\rm C}(\langle v_{ss^\prime}^2\rangle)=
\frac{Z_sZ_{s^\prime}e^2}{3kT_{ss^\prime}},
\ee
where $\langle v_{ss^\prime}^2\rangle= 3kT_{ss^\prime}/\mu_{ss^\prime}$
is the average of the square of the relative velocity (Spitzer~1956).
The averaging of the remaining dependence of the cross section on the
inverse fourth power of the relative velocity then gives, according to
eq.~(\ref{G_n}),
\be
\langle\sigma v\rangle_{ss^\prime}=2\pi a_{ss^\prime}r_{\rm C}^2(a_{ss^\prime})
\ln\left[\frac{\overline{r_{\rm C}}^2+\lambda_{\rm D}^2}
{\overline{r_{\rm C}}^2+r_{\rm min}^2}\right] G_4(\xi),
\label{sv_coul}
\ee
where $a_{ss^\prime}$ is given by eq.~(\ref{defa}) and $G_4(\xi)$ by
eq.~(\ref{g4}).  If the particles are assumed to be pointlike ($r_{\rm
min}=0$), and, in addition, $\lambda_{\rm D}\gg \overline{r_{\rm C}}$
(a condition generally fulfilled in the ISM), we have
\be
\ln\left[1+\left(\frac{\lambda_{\rm D}}{\overline{r_{\rm C}}}\right)^2\right]
\approx 2\ln \left(\frac{\lambda_{\rm D}}{\overline{r_{\rm C}}}\right).
\ee
In this case eq.~(\ref{sv_coul}) for $\xi=0$ gives the standard
expression of the Coulomb momentum transfer rate (Spitzer~1956)
\begin{eqnarray}
\lefteqn{
\langle\sigma v\rangle_{ss^\prime}=
\frac{16\pi^{1/2}Z_s^2Z_{s^\prime}^2 e^4}{3\mu_{ss^\prime}^{1/2}
(2kT_{ss^\prime})^{3/2}}\ln\Lambda_{ss^\prime}
} \nonumber \\
& & \approx 8.48\times 10^{-2} Z_s^2 Z_{s^\prime}^2 
\left(\frac{\mu_{ss^\prime}}{m_H}\right)^{-1/2}
\left(\frac{T}{K}\right)^{-3/2} \ln\Lambda, 
\label{spitz}
\end{eqnarray}
where
\begin{eqnarray}
\lefteqn{\ln\Lambda_{ss^\prime}\equiv \ln\left(\frac{\lambda_{\rm D}}{\overline{r_{\rm C}}}\right)
=\ln\left[\frac{3kT_{ss^\prime}}{Z_sZ_{s^\prime}e^2}
\left(\frac{4\pi e^2}{k}\sum_s\frac{Z_s^2n_s}{T_s}\right)^{-1/2}\right]} \\
& & =9.42-\ln\left\{\frac{Z_sZ_{s^\prime}}{(T_{ss^\prime}/\mbox{K})}
\left[\sum_s\frac{Z_s^2(n_s/\mbox{cm$^{-3}$})}{(T_s/\mbox{K})}\right]^{1/2}\right\}
\label{coul_log}
\end{eqnarray}
is the {\em Coulomb logarithm}.

\subsection{Collisions of ions with charged dust grains}

\begin{figure}
\resizebox{\hsize}{!}{\includegraphics{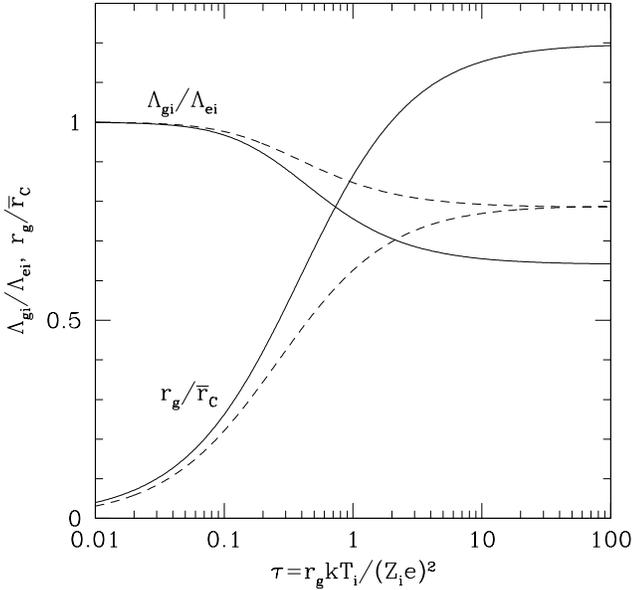}} 
\caption{Argument of the Coulomb logarithm $\Lambda_{gi}$ for
grain-ions interactions normalized to $\Lambda_{ei}$ for pointlike
particles, and ratio $r_g/{\overline r_{\rm C}}$ as a function of the
reduced temperature of the ions $\tau_i=r_g kT_i/(Z_ie)^2$. The {\em
solid} and {\em dashed curves} are for ``light ions'' with effective
atomic weight 1 and ``heavy ions'' with effective atomic weight 25.}
\label{coul}
\end{figure}

In astrophysical applications, eq.~(\ref{spitz}) has often been adopted
to compute the rate of momentum transfer between grain-ion and
grain-electron collisions (Draine \& Salpeter~1979). However, a more
accurate calculation requires a modification of the Coulomb logarithm
to account for the finite size of the grains. Taking $r_{\rm min}=r_g$
in Eq.~(\ref{sv_coul}), and assuming as before $\lambda_{\rm D} \gg
\overline{r_{\rm C}}$, we obtain the modified Coulomb logarithm for
grain-ion (or grain-electron) collisions
\be
\ln\Lambda_{gi}=\ln\left[\frac{\lambda_{\rm D}}{(\overline{r_{\rm C}}^2+r_g^2)^{1/2}}\right],
\ee
which reduces to $\ln(\lambda_{\rm D}/\overline{r_{\rm C}})$ for $r_g=0$,
and approaches $\ln(\lambda_{\rm D}/r_g)$ for $r_g\gg \overline{r_{\rm
C}}$. The latter value of the Coulomb logarithm has been used, e.g., by
Benkadda et al.~(1996) in their study of nonlinear instabilities in
dusty plasmas.  Following Draine \& Sutin~(1987), it is easy to compute the
ratio $r_g/\overline{r_{\rm C}}$ for a variety of astrophysical
conditions, since this quantity depends almost exclusively on the
reduced temperature $\tau_i\equiv r_g kT_i/(Z_ie)^2$,
\be
\frac{r_g}{\overline{r_{\rm C}}}=\frac{3Z_i\tau_i}{Z_g(\tau_i)}.
\ee
Using the analytical approximations for the value of the average grain
charge $Z_ge$ as a function of $\tau_i$ computed by Draine \&
Sutin~(1987), we obtain the modified Coulomb logarithm for grain-ion
collisions given in Fig.~\ref{coul} for ``light'' and ``heavy ions''
(effective atomic weight 1 and 25, respectively). We see from this
figure that the pointlike approximation for the Coulomb logarithm
remains valid for $\tau_i$ less than $\sim 0.1$. For larger values of
$\tau_i$ the value of the Coulomb logarithm is smaller than the
pointlike value, but the effect is a reduction of the constant in
Eq.~(\ref{coul_log}), from 9.42 to 8.97 or 9.18 at worst, for the
``light ion'' and ``heavy ion'' cases, respectively.

For a more general calculation of the Coulomb logarithm for grain-ion
and grain-electron where $r_{\rm min}$ depends on the finite size of
the grain and the relative velocity of the collision, and, in addition,
the approximation $\lambda_{\rm D} \gg \overline{r_{\rm C}}$ is
removed, see Khrapak \& Morfill~(2004).

\section{Analytical approximations}
\label{sec_anal}

In Table~\ref{rates_vd0} we list analytical fitting formulae for the
momentum transfer rate coefficients computed numerically in
Sects.~3--5.  For zero drift velocity, we have approximated the rates
(or their logarithm) with third-order polynomials of the logarithm of
the temperature.  The accuracy of these fitting formulae is $\sim
10$\%.  The dependence on the drift velocity has been approximated,
following GKS and Draine~(1980) by power-laws (or modified power-laws)
fits of the numerical results in terms of the rms velocity, defined as
\be 
v_{\rm rms}\equiv\left(v_d^2+\frac{8kT_{ss^\prime}}{\pi\mu_{ss^\prime}}\right)^{1/2}.
\ee 
These fitting formulae are shown in Table~\ref{rates_vd}.  Within the
range of $v_{\rm rms}$ indicated, the accuracy of the
power-law approximations is better than $\sim 20$\%.

\begin{table}
\caption{Fitting formulae for momentum transfer coefficients as a function
of the gas temperature $T$ (in K) and $\theta\equiv\log (T/\rm{K})$ for $v_d=0$.}
\begin{tabular}{ll}
\hline
species $s,s^\prime$ &  $\langle\sigma v\rangle_{ss^\prime}$ \\
        &   ($10^{-9}$~cm$^3$~s$^{-1}$)         \\
\hline
HCO$^+$,H$_2$  & $T^{1/2}(1.476-1.409\theta+0.555\theta^2-0.0775\theta^3)$ \\
H$_3^+$,H$_2$  & $2.693-1.238\theta+0.664\theta^2-0.089\theta^3$ \\
H$^+$,H$_2$    & $1.003+0.050\theta+0.136\theta^2-0.014\theta^3$ \\
$e$,H$_2$      & $T^{1/2}(0.535+0.203\theta-0.163\theta^2+0.050\theta^3)$ \\
\hline
C$^+$,H    & $1.983+0.425\theta-0.431\theta^2+0.114\theta^3$  \\
H$^+$,H    & $0.649 T^{0.375}$ (GKS) \\
$e$,H      & $T^{1/2}(2.841+0.093\theta+0.245\theta^2-0.089\theta^3)$  \\
\hline
H$^+$,He   & $1.424+7.438\times 10^{-6}T-6.734\times 10^{-9}T^2$ \\
$e$,He     & $0.428 T^{1/2}$ \\
\hline
\end{tabular}
\label{rates_vd0}
\end{table}

\begin{table}
\caption{Fitting formulae for momentum transfer coefficients as function of $v_{\rm rms}$
(in km~s$^{-1}$).}
\begin{tabular}{lll}
\hline
species $s,s^\prime$ &  $\langle\sigma v\rangle_{ss^\prime}$ & $v_{\rm rms}$ \\
        &  (cm$^3$~s$^{-1}$) & (km~s$^{-1}$) \\
\hline
HCO$^+$,H$_2$  & $2.40\times 10^{-9} v_{\rm rms}^{0.6}$  & $0.2\lesssim v_{\rm rms}\lesssim 5$\\
H$_3^+$,H$_2$  & $2.00\times 10^{-9} v_{\rm rms}^{0.15}$ & $1\lesssim v_{\rm rms}\lesssim 10$\\
H$^+$,H$_2$    & $3.89\times 10^{-9} v_{\rm rms}^{-0.02}$ & $1\lesssim v_{\rm rms}\lesssim 10$\\
$e$,H$_2$      & $3.16\times 10^{-11} v_{\rm rms}^{1.3}$ & $20\lesssim v_{\rm rms}\lesssim 200$\\
\hline
C$^+$,H    & $1.74\times 10^{-9} v_{\rm rms}^{0.2}$ & $2\lesssim v_{\rm rms}\lesssim 20$\\
H$^+$,H    & $2.13\times 10^{-9} v_{\rm rms}^{0.75}$ & $v_{\rm rms} \gtrsim 1$ (GKS) \\
$e$,H      & $2.50\times 10^{-10} v_{\rm rms}^{1.2}\exp(-v_{\rm rms}/460)$ & $20\lesssim v_{\rm rms}\lesssim 600$\\
\hline
H$^+$,He   & $1.48\times 10^{-9} v_{\rm rms}^{-0.02}$ & $0.1\lesssim v_{\rm rms} \lesssim 10$  \\
$e$,He     & $7.08\times 10^{-11} v_{\rm rms}$ & $20\lesssim v_{\rm rms} \lesssim 500$  \\
\hline
\end{tabular}
\label{rates_vd}
\end{table}

\section{Conclusions}
\label{sec_concl}

We have derived momentum transfer coefficients for collisions between
ions, electrons, charged dust grains and atomic/molecular hydrogen from
available experimental data and theoretical calculations, within the
classic approach developed by Boltzmann~(1896) and Langevin~(1905). The
numerical results have been approximated with simple analytical
functions of the temperature and the drift velocity between the
colliding species. The main conclusions of our study are the following:

\begin{enumerate}

\item For collisions between molecular ions (HCO$^+$, H$_3^+$) and H$_2$, the
often used polarization approximation valid for an induced electric
dipole attraction is in satisfactory agreement (within
20--50\% of the numerical results). At low temperatures (10--50~K) and
for low values of the drift velocities, the polarization approximation overestimates the actual rate for
collisions between H$^+$ and H$_2$ by a factor of $\sim 3$. In heavily-depleted molecular cloud cores, where H$^+$ may be the dominant ion
(see, e.g., Walmsley et al.~2004), the Langevin approximation should not
be used to compute ambipolar diffusion time scales (see Paper~I).

\item For collisions between electrons and H$_2$, which contribute marginally
to the resistivity of the interstellar gas (see Paper~I), the
polarization approximation fails by orders of magnitude because of the
non-dipolar nature of the interaction, especially at low energies.

\item In the range $10~\mbox{K}\lesssim T \lesssim 10^3~\mbox{K}$,
rate coefficients for collisions between H$^+$ and He atoms are a factor 0.7--1 and 0.1--0.4 of
the corresponding rate for collisions with H$_2$ and H, respectively.
In the same temperature range, the rate coefficient for $e$--He collisions
is a factor 0.4--0.7 of the rate for $e$--H$_2$ collisions and a factor 0.1--0.3 of the rate for $e$--H collisions .

\item Grain-neutral and grain-ion collisions are well represented by hard-spheres and Coulomb interactions, respectively. In the latter
case, slightly different expressions may be adopted for the Coulomb
logarithm, with no significant consequences for applications to ISM
grains.

\end{enumerate}

\acknowledgements
We thank A. Glassgold for helpful discussions and for 
making available to us data on elastic cross sections.
The research of DG is partially supported by the Marie Curie
Research Training networks ``Constellation".

\appendix
\section{Polarization approximation}
\label{sec_pol}

The cross section for the interaction between an ion $i$ of charge $Z_i
e$ and a neutral molecule (or atom) $n$ is determined by the attractive
polarization potential
\be
\label{polpot}
V_{in}(r)=-\frac{p_n Z_se^2}{2r^4},
\ee
where $r$ is the distance between the centers of the ion and the
molecule (much larger than the sizes of the ion and molecule) and
$\alpha_n$ is the polarizability of the molecule. The relative cross
section can be calculated from the classical trajectories allowed by
this potential (see, e.g., Dalgarno, McDowell \& Williams~1958a,b;
Mitchner \& Kruger~1973), and is
\be
\sigma(v_{in})=2.210\pi\left(\frac{p_n Z_ie^2}{\mu_{in} v_{in}^2}\right)^{1/2}.
\ee
In this polarization approximation, the rate coefficient (``Langevin rate'') is 
given by 
\be
\langle\sigma v\rangle_{in}= 2.210\pi\left(\frac{p_n Z_se^2}{\mu}\right)^{1/2},
\ee
which is independent of the gas temperature and the relative drift of
the interacting species.  Osterbrock~(1961) corrected the numerical
coefficient $2.21$ in $2.41$ to account for the repulsive nature of the
interaction potential at small $r$, assuming that for small impact
parameters the scattering is on average isotropic.

The Langevin cross section for collisions of ions with neutrals, taking
into account Osterbrok's (1961) correction, therefore, results in
\be
\sigma_{in}(E_{\rm cm})=2.03\times 10^{-15}Z_i^{1/2}
\left(\frac{p_n}{\mbox{\AA$^3$}}\right)^{1/2}
\left(\frac{E_{\rm cm}}{\rm eV}\right)^{-1/2}~\mbox{cm$^2$},
\label{lange}
\ee
and the corresponding rate is
\be
\langle\sigma v\rangle_{in}= 2.81\times 10^{-9}
Z_i^{1/2}\left(\frac{p_n}{\mbox{\AA$^3$}}\right)^{1/2}
\left(\frac{\mu_{in}}{m_{\rm H}}\right)^{-1/2}~~~\mbox{cm$^3$~s$^{-1}$}.
\label{lanrate}
\ee
Values of the polarizability are $p_{\rm H}=0.667$~\AA$^3$,
$p_{{\rm H}_2}=0.804$~\AA$^3$, and $p_{\rm He}=0.207$ (Osterbrock~1961).

\end{document}